\documentclass[preprint,5p,review,twocolumn]{elsarticle}

\usepackage{amssymb}
\usepackage{amsthm}
\usepackage[utf8]{inputenc}
\usepackage{subfiles}
\usepackage{graphicx}
\usepackage{multirow}
\usepackage{caption}
\usepackage{booktabs}
\usepackage{enumitem}
\usepackage[obeyspaces]{url}
\usepackage{kotex}
\usepackage{lineno,hyperref}
\usepackage{indentfirst}
\usepackage{listings}
\usepackage{amsmath}
\usepackage{makecell}
\usepackage{array}
\usepackage{amsmath}
\usepackage{algorithm}
\usepackage[noend]{algpseudocode}
\usepackage{url}
\usepackage{xurl}
\usepackage{xcolor}

\colorlet{kw}{blue}
\definecolor{com}{rgb}{0,0.6,0.3}

\algrenewcommand{\algorithmiccomment}[1]{\textcolor{kw}{\hfill$\triangleright$\texttt{#1}}}
\algnewcommand{\LeftComment}[1]{\textcolor{kw}{\Statex \quad \(\triangleright\) \texttt{#1}}}

\theoremstyle{definition} 
\newtheorem{observation}{Observation}

\usepackage{kotex}
\usepackage{ulem}

\begin{document}
	
	\begin{frontmatter}

  
		\title{%
			A Method for Decrypting Data Infected with Rhysida Ransomware
		}

		\author[label1,fn1]{Giyoon Kim}
		\ead{gi0412@kookmin.ac.kr}
		
		\author[label1,fn1]{Soojin Kang}
		\ead{szin31@kookmin.ac.kr}
		
		\author[label1,fn1]{Seungjun Baek}
		\ead{hellosj3@kookmin.ac.kr}

        \author[label3]{Kimoon Kim}
		\ead{kkm@kisa.or.kr}
		
		\author[label1,label2]{Jongsung Kim\corref{cor1}}
		\ead{jskim@kookmin.ac.kr}
		\ead[url]{https://dfnc.kookmin.ac.kr}

        \fntext[fn1]{These authors contributed equally to this work.}
  
		\affiliation[label1]{organization={Dept. of Financial Information Security},
			addressline={Kookmin University,77 Jeongneung-Ro, Seongbuk-Gu}, 
			city={Seoul},
			postcode={02707}, 
			country={Republic of Korea}}
		
		\affiliation[label2]{organization={Dept. of Information Security, Cryptology, and Mathematics},
			addressline={Kookmin University,77 Jeongneung-Ro, Seongbuk-Gu}, 
			city={Seoul},
			postcode={02707}, 
			country={Republic of Korea}}

        \affiliation[label3]{
			addressline={Korea Internet \& Security Agency (KISA), 9, Wolgok-gil, Dasi-myeon}, 
			city={Naju},
			postcode={58324}, 
			country={Republic of Korea}}
  
		\cortext[cor1]{Corresponding author.}

		\begin{abstract}
			
               Ransomware is malicious software that is a prominent global cybersecurity threat. Typically, ransomware encrypts data on a system, rendering the victim unable to decrypt it without the attacker’s private key. Subsequently, victims often pay a substantial ransom to recover their data, yet some may still incur damage or loss. This study examines Rhysida ransomware, which caused significant damage in the second half of 2023, and proposes a decryption method. Rhysida ransomware employed a secure random number generator to generate the encryption key and subsequently encrypt the data. However, an implementation vulnerability existed that enabled us to regenerate the internal state of the random number generator at the time of infection.
               We successfully decrypted the data using the regenerated random number generator.
To the best of our knowledge, this is the first successful decryption of Rhysida ransomware. We aspire for our work to contribute to mitigating the damage inflicted by the Rhysida ransomware.

		\end{abstract}


		
	\end{frontmatter}
	
	\renewcommand\linenumberfont{\normalfont\tiny\sffamily\color{gray}}

	\section{Introduction}
	\label{intro}
	
The world is undergoing rapid digitalization, with digital technology services gaining unprecedented popularity. Concurrently, the rise in cybersecurity threats is noteworthy.  Ransomware is a prominent threat that infiltrates systems to encrypt user data and demand payment for decryption. Ransomware, through data encryption, not only leads to data loss but also disrupts the organization’s services. When ransomware infects an infrastructure that serves  a large user base, both the infrastructure and its users suffer damage. 
As a result, for the victim, paying the ransom emerges as a practical choice.

Advanced ransomware not only encrypts data but also carries out data exfiltration, frequently using double extortion tactics that threaten to delete the exfiltrated data.
Furthermore, there has been a rise in triple extortion schemes, offering analysis of victims’ vulnerabilities and advice on prevention~\cite{acer2021}, As the scale of ransomware damage grows, global recommendations for proactive responses are actively disseminated and formulated into policies. Implementing proactive measures such as installing anti-virus software, regular software updates, and creating backups can effectively minimize the damage caused by ransomware. Despite these efforts, ransomware-induced damage persists, underscoring the significance of post-counteraction studies for data after ransomware infection.


When analyzing ransomware, the most desirable thing is to find the key used to encrypt the data, but this is not an easy task.
Since ransomware typically employs a hybrid encryption system, it is difficult to find the key without obtaining the attacker's private key.
Therefore, researchers diligently work to develop decryption methods through reverse engineering by identifying weaknesses in ransomware implementations.

In developing ransomware decryption methods, there are instances where key regeneration is possible.
If key regeneration succeeds, the next step is to determine the specific key used to encrypt each encrypted file.
Ransomware often employs multiprocessing for swift encryption, thereby introducing a potential randomization factor in the order of file encryption, despite the sequential generation of encryption keys.
This can pose a challenge for researchers developing decryption tools.

In this study, we analyze the Rhysida ransomware, which emerged in 2023 and affected numerous victims, including hospitals. Despite its randomized order of file access for encryption and the absence of a developed decryption tool, our results reveal the feasibility of decrypting Rhysida ransomware.

	\paragraph{\textbf{Our Contributions}}
	Our contributions are summarized as follows:
	\begin{enumerate}
     
        \item We performed an in-depth analysis of Rhysida Ransomware.
       Through a comprehensive analysis of Rhysida Ransomware, we identified an implementation vulnerability, enabling us to regenerate the encryption key used by the malware. Subsequently, we developed a recovery tool\footnote{Currently, the developed recovery tool is being distributed through the Korea Internet \& Security Agency (KISA)~\cite{kisarhysida}} for systems infected with Rhysida ransomware, which requires no additional information.

	\end{enumerate}
 
	\paragraph{\textbf{Paper Organization}}

This paper is organized as follows.  Section~\ref{sectionRelatedWork}  provides an introduction to related work. Section~\ref{sectionRhysidaAnalysis]} outlines the operation of Rhysida ransomware. Section~\ref{sectionRhysidaDecrypt} details a decryption method for Rhysida ransomware derived from our analysis. 
The paper concludes with Section~\ref{sectionCon}.
 
	\section{Related Work}
	\label{sectionRelatedWork}

    Ransomware research spans various fields and is categorized into proactive and reactive responses. Proactive responses leverage domain knowledge to minimize ransomware damage by employing techniques such as hash databases, Yara rules, and similarity analysis. Prevention methods, such as creating decoys, halt infection onset and minimize harm. Reactive responses focus on damage recovery after infection, involving tasks such as collecting encryption keys, backing up data, and developing decryption methods. Similar to this study, other some studies attempted to decrypt data by identifying vulnerabilities in ransomware.

    Lee et al. analyzed Magniber v2 ransomware and devised a method to decrypt it without the need for the attacker’s private key~\cite{lee2021magniber}.
    They identified a vulnerability in the pseudo-random number generator used to generate encryption keys in Magniber v2, which significantly reduced the number of potential encryption keys. Employing CBC padding rules and NIST’s SP800-22 randomness test suite~\cite{rukhin2001nist}, they successfully determined the actual encryption key, enabling the decryption of data.
    In addition, Lee et al. analyzed the encryption processes of five major ransomware variants in 2019 (Gandcrab v5, Clop, Sodinokibi, Phobos, and LooCipher), outlining scenarios where decryption is possible~\cite{lee2019study}. LooCipher exhibited a vulnerability in its pseudo-random number generator, as identified by Lee, who demonstrated the possibility of decrypting all data by exploiting this weakness. 
     
    Kang et al. analyzed Ragnar Locker ransomware and proposed a data decryption method~\cite{kang2021study}.
    Their research revealed that the Rangnar Locker improperly used stream ciphers, making it susceptible to reused key attacks. 
    They concluded that if the originals of certain data could be obtained, it was possible to decrypt all remaining data without the originals through a key reuse attack. 
     
    Yuste et al. developed a decryption method for the Avaddon ransomware, developing an open source tool for decrypting the Avaddon ransomware family~\cite{yuste2021avaddon}.  They further integrated their tool with existing antivirus engines to decrypt ransomware using an encryption process similar to that of Avaddon. 

    Kim et al. analyzed the HIVE ransomware and developed a method to decrypt encrypted data without requiring the attacker’s private key~\cite{kim2022method}.
    They demonstrated that the HIVE ransomware uses a custom encryption algorithm with vulnerabilities. 
    The ransomware encrypts the original data by XORing it with two random numbers, thereby generating a linear equation.
    The authors demonstrated that this equation could be used to infer the encryption key and showed the possibility that if the key is reused, the encrypted data can be decrypted.
     
    These studies achieved data decryption by exploiting implementation vulnerabilities in ransomware. Although there are similarities with our study, the findings are not applicable to the Rhysida ransomware that we analyzed. In addition, each of these studies had a clear order in which files were encrypted. 
    In contrast, our study distinguishes itself by assuming an unknown order of file access for encryption.

	\section{Analysis of Rhysida Ransomware}
	\label{sectionRhysidaAnalysis]}

    In this section, we analyze the encryption process of Rhysida ransomware. 
    The Rhysida ransomware, compiled with MinGW GNU version 6.3.0, exists as either a 32-bit or 64-bit PE file.
    We conduct our analysis on 11 Rhysida ransomware sample files, and Table~\ref{tab:sample_properties1} displays the SHA256 hash values for each sample. 
    Using the x64 debugger~\cite{x64dbg}, we examined the ransomware. 
    Rhysida ransomware exclusively uses LibTomCrypt~\cite{libtomcrypt} for encryption. To expedite data encryption, Rhysida ransomware performs parallel processing by creating sub-threads equivalent to the number of processors on the victim PC.
    This chapter describes Rhysida ransomware, with a focus on three essential parts for decrypting the data: identification of factors for regenerating the encryption key, determination of the encryption target, and the data encryption process.

	\begin{table}[ht!]
		\centering
		\caption{SHA256 hash values of the Rhysida ransomware sample files}
		\label{tab:sample_properties1}
			\resizebox{1\columnwidth}{!}{
				\begin{tabular}{c}
					\toprule
					SHA256 Hash Values \\ \midrule
					 
					\texttt{A864282FEA5A536510AE86C77CE46F7827687783628E4F2CEB5BF2C41B8CD3C6} \\
					\texttt{0BB0E1FCFF8CCF54C6F9ECFD4BBB6757F6A25CB0E7A173D12CF0F402A3AE706F} \\
					\texttt{F6F74E05E24DD2E4E60E5FB50F73FC720EE826A43F2F0056E5B88724FA06FBAB} \\
					\texttt{1A9C27E5BE8C58DA1C02FC4245A07831D5D431CDD1A91CD35D2DD0AD62DA71CD} \\
					\texttt{2C5D3FEA7AD3C9C49E9C1A154370229C86C48FBAF7044213FD85D31EFCEBF7F6} \\
					\texttt{3D2013C2BA0AA1C0475CAB186DDF3D9005133FE5F88B5D8604B46673B96A40D8} \\
					\texttt{67A78B39E760E3460A135A7E4FA096AB6CE6B013658103890C866D9401928BA5} \\
					\texttt{250E81EEB4DF4649CCB13E271AE3F80D44995B2F8FFCA7A2C5E1C738546C2AB1} \\
					\texttt{258DDD78655AC0587F64D7146E52549115B67465302C0CBD15A0CBA746F05595} \\
					\texttt{3518195C256AA940C607F8534C91B5A9CD453C7417810DE3CD4D262E2906D24F} \\
					\texttt{D5C2F87033A5BAEEB1B5B681F2C4A156FF1C05CCD1BFDAF6EAE019FC4D5320EE} \\
					\bottomrule
				\end{tabular}
			}
		\end{table}

        \subsection{Identification the factors for regenerating the encryption key}
		\label{subSectionReGenKey}
		
  Decrypting data encrypted using a symmetric-key cryptographic algorithm requires the encryption key used in the process.
  Since encryption keys can be generated in various methods, it is important to identify the factors used by ransomware in the key generation process during data encryption.
  
  In this study, we categorize encryption key generation methods into two types: generation based on a specific value (not a random number) or generation based on a random number. 
  The former involves generating the encryption key using a fixed or inferred specific value, which is easily regenerable through ransomware analysis.
  In the latter case, the key is generated based on a random number, necessitating the identification of the random number generator used for encryption and the internal state of the generator at that time. 
  The random number generator takes a seed as input, sets it as the initial internal state, and generates a sequence of random numbers according to a defined rule.
  Therefore, if we can identify the initial internal state, regenerating the random number becomes feasible.

\begin{figure*}[bht!]
    \centering
    \includegraphics[width=\linewidth]{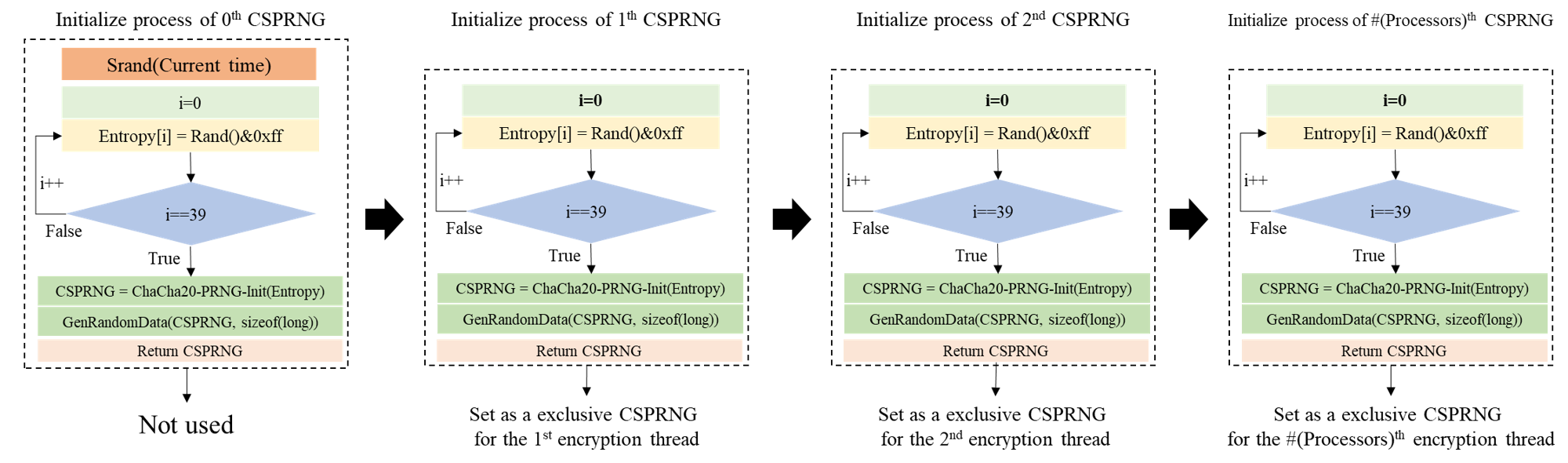}
    \caption{CSPRNG initialize process}
    \label{fig:initcsprng}
\end{figure*}

Rhysida ransomware uses a cryptographically secure pseudo-random number generator (CSPRNG) to generate the encryption key. 
This generator uses a cryptographically secure algorithm to generate random numbers.
Our analysis revealed  that Rhysida ransomware uses CSPRNG, which is based on the ChaCha20 algorithm provided by the LibTomCrypt library. 
To enhance the randomness of its internal state, CSPRNG can be input with random entropy data. 
Rhysida ransomware updates its internal state once by incorporating specific entropy data before using CSPRNG. 
The entropy data is a 40--byte random number generated by the rand function in the C standard library. 
The initial \textit{seed} of the \textit{rand} function is set based on the time at which Rhysida ransomware is running. 
After updating the internal state of CSPRNG, Rhysida ransomware verifies its proper functioning by generating a random number equivalent to \textit{sizeof(long)}.
During this process, the internal state of CSPRNG is incremented by 4 bytes for x86 and 8 bytes for x64, depending on the compiled version of Rhysida ransomware. 
The initialization of the CSPRNG process terminates if the random number is successfully generated. 
In addition, to parallelize data encryption, Rhysida ransomware creates sub-threads equal to the number of processors on the victim PC, assigning a distinct CSPRNG to each thread. 
However, in contrast to sub-threads, which are created corresponding to the number of processes, the initialization of the CSPRNG process by Rhysida ransomware is performed for a total of \#(Processors) + 1. 
Consequently, this leads to the generation of \#(Processors) + 1 CSPRNGs.
However, the first initialized CSPRNG remains unused, and from the subsequent CSPRNG onward, it is sequentially assigned to previously created sub-threads.  
Figure~\ref{fig:initcsprng} shows the initialization process of the Rhysida ransomware’s random number generator.

		\begin{observation}
			\label{obsCSPRNG}
Our analysis reveals that the random number generated by the CSPRNG is based on the execution time of the Rhysida ransomware. 
The time value used as a \textit{seed}, being 32-bit data, does not offer a large space for conducting an exhaustive search
Thus, the number of possible cases of CSPRNG is up to $2^{32}$.
		\end{observation}
	
    \subsection{Determination of the encryption target}
		
 Before encrypting data, ransomware selects a target to be encrypted. 
 The process of identifying encryption targets can be categorized into two methods. 
 The first method is the on-the-fly approach, which involves traversing folders and encrypting files immediately. 
 The second method is the list-up approach, which involves traversing the folder, initially listing the files, and subsequently performing encryption by referencing these listed files. 
 Traversing folders encompasses four types of approaches, including depth-first search (DFS), which prioritizes entering a folder first, and breadth-first search (BFS), which traverses files within a folder first. 
 In each case, ascending and descending orders can be applied.
 When the entire process of identifying encryption targets is exclusively performed by the main process, determining the order of file encryption becomes feasible through analysis. 
 However, if this process is executed in parallel, it can pose challenges for an analyst in determining the order of file encryption.

\begin{figure}[ht!]
    \centering
    \includegraphics[width=1\columnwidth]{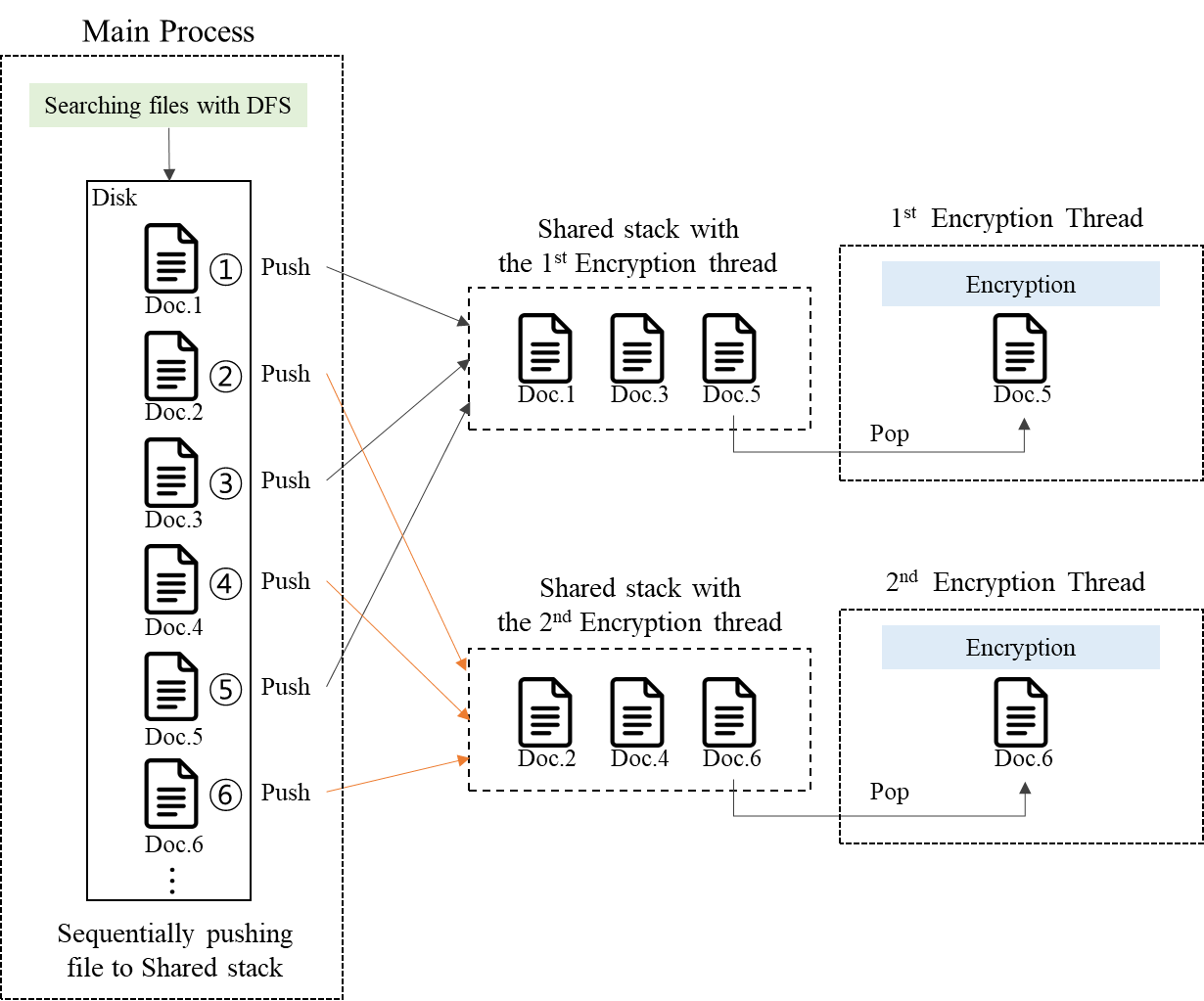}
    \caption{Rhysida ransomware file selection process when there are two processes on the PC}
    \label{fig:listup}
\end{figure}

 Our findings indicate that the main process of Rhysida ransomware compiles a list of encryption targets, and the encryption threads reference this list for data encryption. 
 The main process traverses the entire drive in ascending order, employing DFS, and subsequently pushes the encryption target list to the stack of encryption threads. 
  Simultaneously, the encryption thread performs the encryption by popping the target to be encrypted from its own stack, with the order of file access for encryption being determined by the last in, first out (LIFO) order.
 If the current stack reaches full capacity, the main process suspends pushing data and waits until the current stack is ready to accommodate data. 
Due to the distinct threads handling the push and pop processes, the final order of the encrypted files becomes disordered. 
Figure~\ref{fig:listup} outlines the file selection process used for encryption by Rhysida ransomware.

    To successfully decrypt data after regenerating the encryption key, it is crucial to identify the order in which the encryption keys were used.
   However, in the absence of explicit rules governing the order of file encryption, a workaround becomes necessary. 
   The file system stores \textit{mtime} (modification time) for each file. When a file is encrypted, its \textit{mtime} is changed. 
   \textit{This change provides insight into the order in which the file was encrypted. However, in cases where files share the same \textit{mtime} (due to small file size or intermittent encryption for swift processing), distinguishing the order of file access for encryption becomes essential.}
   This leads to the following observation.
   
		\begin{observation}
			\label{obsGenFileList}
    Accurately determining the whole order of files for encryption is possible when distinguishing the order in instances where the files have the same \textit{mtime}.

		\end{observation}

  \subsection{Data encryption process}
	
  Ransomware commonly encrypts data in two ways. The first method is the \textit{whole} encryption technique, which encrypts the entire file. The second method is the \textit{intermittent} encryption technique, which gained popularity with the emergence of LockBit ransomware~\cite{intermittentEncryption}. Intermittent encryption involves partially encrypting files and can be implemented in various ways. 
  A typical approach is to encrypt \textit{n}bytes and then leave \textit{m}bytes unencrypted, repeating this process until the entire file is encrypted. 
  While parts of the file remain in plaintext due to intermittent encryption, the file becomes unusable, and the encryption speed increases.

Rhysida ransomware implemented intermittent encryption, encrypting 1 MiB from a specific offset (\textit{O}). The calculation of \textit{O}is as follows:

		\begin{align}
			div &= \begin{cases}
				4 & \text{if} \quad \text{fileSize} \geq 4\text{--MiB} \\
				\lfloor \frac{\text{fileSize}}{1 \text{ MiB}} \rfloor & \text{elif } 1\text{--MiB} \leq \text{fileSize} <4 \text{--MiB} \\
				1 & \text{else}
				\end{cases} \\
			O &= \{\lfloor \frac{\text{fileSize}}{div} \rfloor \times i \mid i \in \text{range(div)}\}	
		\end{align}
		
If the fileSize is less than 1 MiB, the entire data is encrypted. 
For the fileSize is ranging from 1 MiB to less than 2 MiB, the top 1 MiB is encrypted. 
In the case of fileSize ranging from 2 MiB to less than 3 MiB, the data are divided into two parts, and the starting 1 MiB of each part is encrypted. 
For the fileSize ranging from 3 MiB to less than 4 MiB, the data are divided into three parts, and the starting 1 MiB of each part is encrypted. 
For the fileSize is 4 MiB or more, the data are divided into four parts, and the starting 1 MiB of each part is encrypted.

 The data was encrypted using the AES--256--CTR (Little-Endian) encryption mode, utilizing a 32--byte encryption key and a 16--byte IV. 
 These key and IV were generated by the CSPRNG assigned to each encryption thread.
  Rhysida ransomware concludes the process by using a hardcoded RSA--4096 public key to encrypt each randomly generated encryption key and IV, storing them at the end of the file.

\begin{figure*}[htb!]
            \centering
            \includegraphics[width=0.95\textwidth]{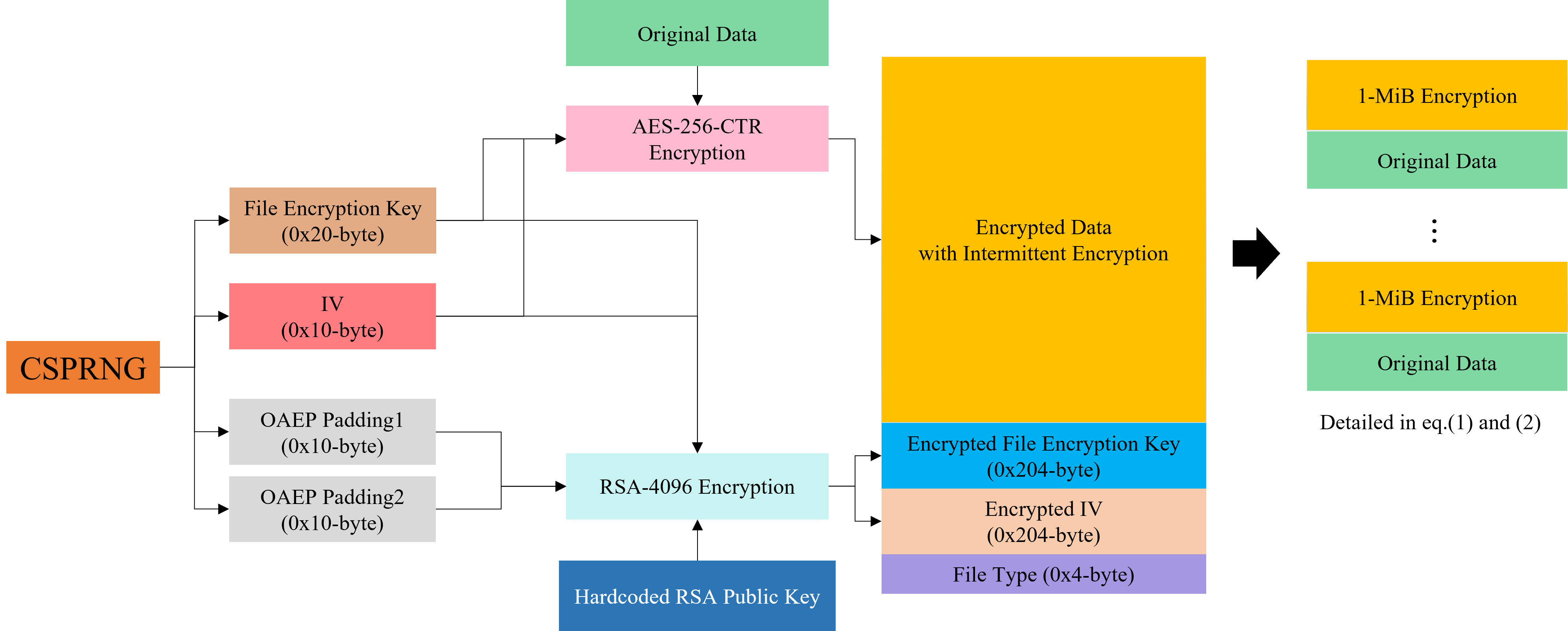}
            \caption{The intermittent encryption process of Rhysida ransomware}
            \label{fig:encryption_process}
        \end{figure*}

Both AES and RSA, which were used for encryption, employed the LibTomCrypt library. For RSA, Rhysida ransomware uses the optimal asymmetric encryption padding (OAEP) technique. OAEP requires a random number and a hash function. Therefore, the Rhysida ransomware uses a CSPRNG and a block cipher-based hash function assigned to the encryption thread. Figure~\ref{fig:encryption_process} outlines the data encryption process of the Rhysida ransomware.

    Successfully decrypting data encrypted by ransomware requires precise identification of both the encryption method and the encryption key.
   Since the encryption method of Rhysida ransomware is already known, it does not need to be considered.
   Indeed, it is important to determine the encryption key from the random numbers generated by the CSPRNG.
   The CSPRNG generates random numbers four times during the encryption process: for encryption key generation, IV generation, encryption of the encryption key (RSA), and encryption of the IV (RSA).
    The encryption key is 32 bytes, the IV is 16 bytes, and the size of the random number generated within the OAEP is 16 bytes. Within the OAEP, two instances of 16-byte random numbers are generated: one during the encryption of the encryption key and another during the encryption of the IV.
    This leads to the following observation.
   
		\begin{observation}
			\label{obsDataEncryption}
    In the encryption process of the Rhysida ransomware, the encryption thread generates 80 bytes of random numbers when encrypting a single file. Of these, the first 48 bytes are used as the encryption key and the IV.
		\end{observation}

		\section{Decryption Methodology of Rhysida Ransomware}
		\label{sectionRhysidaDecrypt}
		
	We analyzed the Rhysida ransomware operation described in the previous sections to identify the elements required for decryption. Decrypting data encrypted by Rhysida ransomware requires reconstruction of the encryption key and determination of the order of file encryption. This chapter provides a detailed description of each process and presents the results of decrypting the Rhysida ransomware based on these findings.

  \subsection{Methods for regenerating the encryption keys}
		
		The initially used \textit{seed }must regenerate all the correct encryption keys. With $2^{32}$ cases of \textit{seed}, there are enough possibilities to regenerate all types of encryption keys. However, this approach is inefficient, requiring more than $2^{32}$ decryption trials for each file. Therefore, a validation method is needed to find the \textit{seed}. As a verification method, we decrypted the first file encrypted by each encryption thread.

      Initially, we assessed the number of encryption threads (processors) employed on the victim’s system, which is represented as \#(Processors). Following this, we established a random \textit{seed }to generate a CSPRNG in a manner similar to that of the Rhysida ransomware. Subsequently, we arranged the files encrypted by the Rhysida ransomware in ascending order based on \textit{mtime }and selected the first \#(Processors). In this context, we evaluated whether the number of files that overlapped with the \textit{mtime} of the first file is greater than or less than \#(Processors). The following conditions were considered:

        \begin{enumerate}[label=\roman*]
        \item If the number of files that overlap with the \textit{mtime} of the initial file is less than \#(Processors).

Rhysida ransomware prevents overlap when the main process assigns the encryption target to the encryption thread. Subsequently, the encryption key is generated sequentially from the CSPRNG to encrypt the data. Therefore, in this case, the selected \#(Processors) files must include at least one file encrypted with the first encryption key per encryption thread.

        \item The number of files that overlap with the first file's \textit{mtime} is more than \#(Processors).
        
      In cases where the file size is sufficiently small, leading to rapid encryption, the \textit{mtime} overlap may be greater than \#(Processors). Consequently, \#(Processors) files may not include any files encrypted with the first encryption key per encryption thread. For example, if the second or third file encrypted by a thread shares the same \textit{mtime} as the first file, the order of file encryption may vary depending on the sorting method. This implies that the number of files to be selected to ensure that at least one file is encrypted with the first encryption key is \#(Processors)$\times$\textit{N}, where \textit{N} is the smallest integer that satisfies (\#(Processors)$\times$\textit{N $>$ mtime }is the number of overlapping files).

		\end{enumerate}

    \begin{figure*}[bht!]
        \centering
        \includegraphics[width=0.95\textwidth]{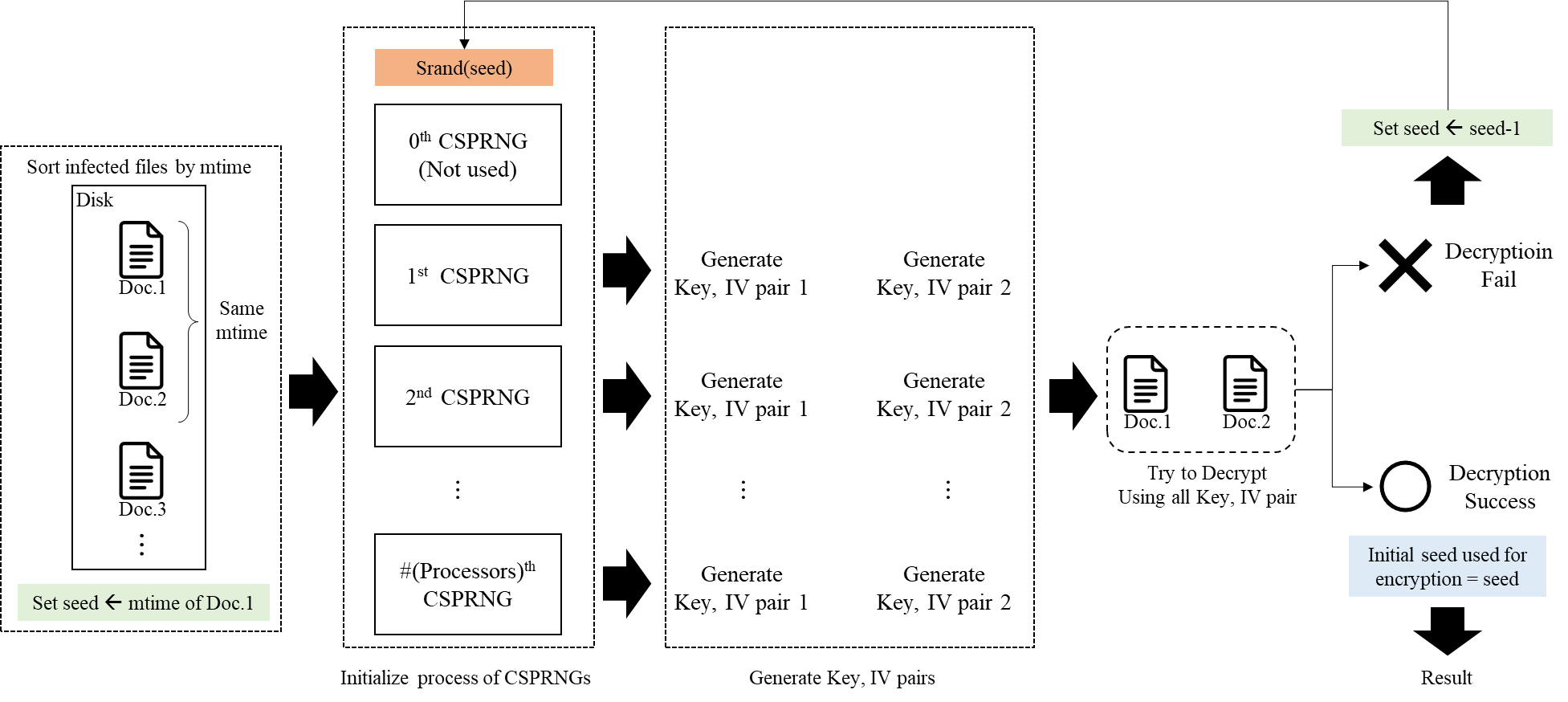}
        \caption{The process of obtaining the initial \textit{seed} of the Rhysida ransomware}
        \label{fig:fig_finding_seed}
    \end{figure*}
  
         Finally, we decrypted the selected \#(Processors)$\times$\textit{N}  files with the first encryption key per encryption thread.

      Correct decryption of encrypted data relies on the encryption key used at the time of encryption. Therefore, if the correct encryption key is regenerated, one or more files will be decrypted correctly. By iteratively generating encryption keys with different \textit{seeds} and identifying a file decrypted correctly using the key, we can conclude that we have found the initial \textit{seed}. Furthermore, through analysis, the initial \textit{seed} used by Rhysida ransomware is the time at the moment of encryption. Efficient retrieval of the initial \textit{seed} involves examining integers smaller than \textit{mtime }of files encrypted by Rhysida ransomware. Figure~\ref{fig:fig_finding_seed}  outlines the process of obtaining the initial \textit{seed }for decrypting Rhysida ransomware.

  \subsection{Methods for reconstructing the order of file encryption}
  
		 The file list generation process of the Rhysida ransomware can be inferred from the operation of the main process and the \textit{mtime} of the files.  The main process traverses drives based on the DFS in ascending order. file list generation process of the Rhysida ransomware can be inferred from the operation of the main process and the \textit{mtime }of the files. The main process traverses drives based on the DFS in ascending order. In addition, during drive traversal, it stores the \textit{j}--th file in the \textit{stack }in the (\textit{j }mod \#(Processors))+1--th encryption thread.  
		
  While the actual order of encryption may not be sequential due to LIFO, we can assert that the list being assigned to the encryption thread is even. Replicating this process allows us to compile a list of files accessed by each thread. However, as mentioned above, the order of encryption within the list may not be sequential.
  Therefore, it is essential to rearrange the order of file for encryption in list.

            Sorting the list of files accessed by threads based on \textit{mtime} allows us to determine the comprehensive file access order. First, we assume that the \textit{mtime} values of the encrypted files are all different. For a specific thread engaged in encryption to proceed with encrypting the subsequent file in the list, it must complete the encryption of the current file. 
            Given distinct \textit{mtime} values, the list of files sorted by \textit{mtime} represents the order of file encryption.

       Subsequently, we consider scenarios where some files share identical \textit{mtime}. If the size of the file to be encrypted is very small, encryption may be completed in the same \textit{mtime}. In such instances, the list of files can be divided into a subset with identical \textit{mtime} and other files. 
       The order of file encryption is guaranteed for all files except the subset. 
       For example, if there are five file lists, where the first file shares the same \textit{mtime} as the second file and the fourth file shares \textit{mtime} with the fifth file, encryption is performed sequentially.
       Consequently, the first and second files are encrypted using either the first or second encryption key. The third file, which has a different \textit{mtime} from the second and fourth files, is encrypted with the third encryption key. In addition, it can be determined that the fourth and fifth files are encrypted using the fourth or fifth encryption key, respectively. 
       In essence, the order of file encryption can be inferred by resolving the subset resulting from the same \textit{mtime}.

    Before deducing the order of file encryption, we must consider the maximum number of possible overlaps of \textit{mtime}.
    In the Windows environment, \textit{mtime }is calculated in \textit{µ}s. Consequently, all encryptable files within a 1 \textit{µ}s timeframe share the same \textit{mtime}. For AES in the LibTomCrypt library used by the Rhysida ransomware, the AES roundkey generation process requires approximately 1400 cycles, and the encryption process requires 18 cycles per byte.  
     Therefore, with a CPU performance of 6 GHz and no additional processing cycles, such as I/O, except for encryption, up to four data can be encrypted in the same \textit{mtime}. However, in this case, the combined size of the four data cannot exceed 22 bytes. If three data are to be encrypted, their total size cannot exceed 100 bytes. Furthermore, for two data to be encrypted, the combined size must not exceed 177 bytes. Consequently, we determined that having up to four data with the same \textit{mtime} is impractical. In these cases, it is effective to decrypt all files in a subset using all encryption keys.

  \subsection{Methods for decrypting encrypted data}

\begin{figure}[ht!]
            \centering
            \includegraphics[width=0.8\columnwidth]{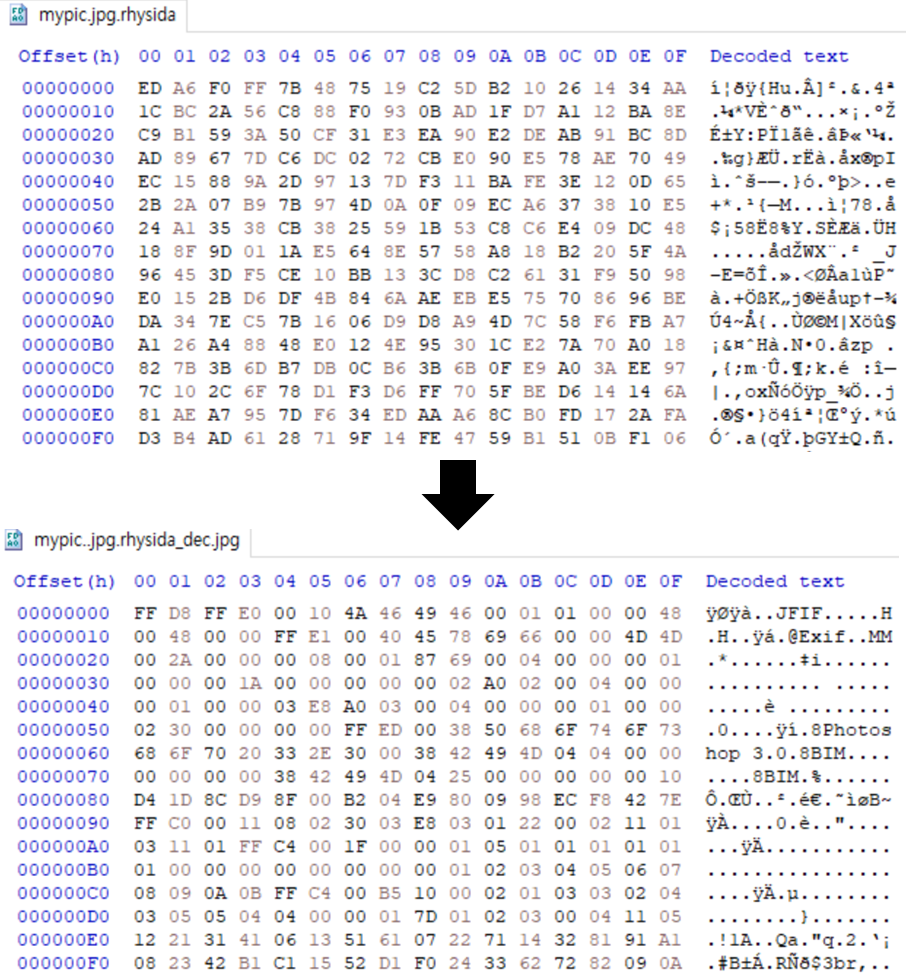}
            \caption{Decryption result of Rhysida ransomware}
            \label{fig:rhysidec}
        \end{figure}

        Rhysida ransomware used intermittent encryption with the AES-–256--CTR algorithm. Data encrypted in CTR mode can be decrypted by performing encryption once more., effectively constituting a decryption process. However, upon analyzing the encryption procedure of Rhysida ransomware, we observed an increase in file size during encryption compared with the original. To facilitate the normal decryption of data upon ransom payment, Rhysida ransomware appends the following information at the end of the file: RSA--encrypted encryption key, RSA--encrypted IV and 4--byte version information. Consequently, the encrypted data are larger than the original data by 0x40C bytes. Because intermittent encryption is based on the size of the original file, it is crucial to eliminate these appended data. Otherwise, the calculation of the encryption location will be incorrect. Consequently, to ensure accurate decryption, it is imperative to delete the last 0x40C bytes of encrypted data before performing intermittent encryption with the correct encryption key. Figure~\ref{fig:rhysidec} shows the decryption results for files encrypted by Rhysida ransomware. Our results demonstrate that encrypted files can be successfully decrypted.

		\section{Conclusion}
		\label{sectionCon}
        
            This paper introduces a method for decrypting files affected by Rhysida ransomware through a thorough analysis that uncovers implementation vulnerabilities. Leveraging these vulnerabilities, we successfully reconstructed the encryption key and restored the encrypted system. 
            Despite the prevailing belief that ransomware renders data irretrievable without paying the ransom. Although these studies have a limited scope, it is important to acknowledge that certain ransomwares, as exemplified in this paper, can be successfully decrypted. We anticipate further studies in this direction to aid ransomware victims, and that our findings will benefit those affected by the Rhysida ransomware.

\section*{Acknowledgement}
\label{ack}

This work was supported by Korea Internet \& Security Agency (KISA) grant funded by the Korea government (Ministry of Science and ICT, MSIT)

		\bibliographystyle{elsarticle-harv}
		\bibliography{bibilography}
	\end{document}